\documentclass{ws-procs975x65}            
\begin{document}
\title{Neutrino interaction with background matter in a noninertial frame}

\author{Maxim Dvornikov$^*$}

\address{Physics Faculty, National Research Tomsk State University,
36 Lenin Ave., 634050 Tomsk, Russia;\\
Pushkov Institute of Terrestrial Magnetism, Ionosphere
and Radiowave Propagation (IZMIRAN),
142190 Troitsk, Moscow, Russia;\\
Institute of Physics, University of S\~{a}o Paulo,
CP 66318, CEP 05315-970 S\~{a}o Paulo, SP, Brazil\\
$^*$E-mail: maxdvo@izmiran.ru}

\begin{abstract}
We study Dirac neutrinos propagating in
rotating background matter. First we derive
the Dirac equation for a single massive neutrino in the noninertial frame,
where matter is at rest. This equation is written in the effective curved space-time corresponding to the corotating frame. We find the exact solution of the Dirac equation. The neutrino energy levels for ultrarelativistic particles are obtained. Then we
discuss several neutrino mass eigenstates, with a nonzero mixing between them, interacting with rotating background matter.
We derive the effective
Schr\"{o}dinger equation governing neutrino flavor oscillations
in rotating matter. The new resonance condition for neutrino
oscillations is obtained. We also examine the correction to the resonance condition caused by the matter rotation.
\end{abstract}


\bodymatter

\section{Introduction}

Nowadays it is commonly believed that neutrinos are massive particles
and there is a nonzero mixing between different neutrino generations.
These neutrino properties result in transitions between neutrino flavors,
or neutrino oscillations. It is also known that various
external fields, like electroweak interaction of neutrinos with background
fermions, neutrino electromagnetic interaction,
and neutrino interaction with a strong gravitational field,
can also influence the process of neutrino oscillations.

As shown in Ref.~\citenum{Lam01}, noninertial effects in accelerated
and rotating frames can affect neutrino propagation and oscillations.
The consideration of the reference frame rotation is particularly
important for astrophysical neutrinos emitted by a rapidly rotating
compact star, e.g., a pulsar. For example, the possibility of the pulsar spin down
by the neutrino emission and interaction with rotating matter was
recently discussed in Ref.~\citenum{DvoDib10}. It should be noted that besides elementary particle physics, various processes in noninertial frames are actively studied in condensed matter physics. For instance, the enhancement of the spin current in a semiconductor moving with an acceleration was recently predicted in Ref.~\citenum{BasCho13}.

In the present work we summarize the results of Ref.~\citenum{Dvo14}, where the neutrino interaction with background
matter in a rotating frame was studied. We assume that neutrinos can interact
with background fermions by means of electroweak forces. We also take
that neutrino mass eigenstates are Dirac particles. In our treatment
we account for the noninertial effects since we find the exact solution
of a Dirac equation for a neutrino moving in the curved space-time with
a metric corresponding a rotating frame.

This work is organized in the following way. In Sec.~\ref{sec:NUMATFLS},
we start with the brief description of the neutrino interaction with
background matter in Minkowski space-time. We consider matter moving
with a constant velocity and discuss both neutrino flavor and mass eigenstates.
Then, in Sec.~\ref{sec:MASSNUNONIF}, we study background matter
moving with an acceleration. The Dirac equation, including noninertial
effects, for a neutrino interacting with such matter is written down
in a frame where matter is at rest. In Sec.~\ref{sub:ROTATION},
we solve the Dirac equation and find the neutrino energy spectrum
for ultrarelativistic neutrinos moving in matter rotating with a constant
angular velocity. Then, in Sec.~\ref{sec:OSC}, we apply our results
for the description of neutrino oscillations in rotating background
matter. The effective Schr\"{o}dinger equation governing neutrino oscillations
is derived and the new resonance condition is obtained. We consider
how the matter rotation can affect the resonance in neutrino oscillation
in a realistic astrophysical situation. Finally, in Sec.~\ref{sec:SUMMARY},
we summarize our results.

\section{Neutrino interaction with background matter\label{sec:NUMATFLS}}

In this section we describe the interaction of different neutrino
flavors with background matter in a flat space-time. We discuss a
general case of matter moving with a constant mean velocity and having
a mean polarization. Then we consider the matter interaction of neutrino
mass eigenstates, which are supposed to be Dirac particles.

The interaction of the neutrino flavor eigenstates $\nu_{\alpha}$,
$\alpha=e,\mu,\tau$, with background matter in the flat space-time
is described by the following effective Lagrangian\cite{GiuKim07}:
\begin{equation}\label{eq:Lf}
  \mathcal{L}_{\mathrm{eff}} =
  -\sum_{\alpha}\bar{\nu}_{\alpha}\gamma_{\mu}^{\mathrm{L}}\nu_{\alpha}\cdot f_{\alpha}^{\mu},
\end{equation}
where $\gamma_{\mu}^{\mathrm{L}}=\gamma_{\mu}(1-\gamma^{5})/2$, $\gamma^{\mu}=(\gamma^{0},\bm{\gamma})$
are the Dirac matrices, and $\gamma^{5}=\mathrm{i}\gamma^{0}\gamma^{1}\gamma^{2}\gamma^{3}$.

The interaction Lagrangian in Eq.~(\ref{eq:Lf}) is derived in the
mean field approximation using the effective external currents $f_{\alpha}^{\mu}$
depending on the characteristics of background matter as\cite{DvoStu02}
\begin{equation}\label{eq:flavnueffp}
  f_{\alpha}^{\mu} =
  \sqrt{2}G_{\mathrm{F}}\sum_{f}
  \left(
    q_{\alpha,f}^{(1)}j_{f}^{\mu}+q_{\alpha,f}^{(2)}\lambda_{f}^{\mu}
  \right),
\end{equation}
where $G_{\mathrm{F}}$ is the Fermi constant and the sum is taken
over all background fermions $f$. Here
\begin{equation}\label{eq:jms}
  j_{f}^{\mu}=n_{f}u_{f}^{\mu},
\end{equation}
is the hydrodynamic current and
\begin{equation}\label{eq:lambdams}
  \lambda_{f}^{\mu} =
  n_{f}
  \left(
    (\bm{\zeta}_{f}\mathbf{u}_{f}),
    \bm{\zeta}_{f}+\frac{\mathbf{u}_{f}(\bm{\zeta}_{f}\mathbf{u}_{f})}{1+u_{f}^{0}}
  \right),
\end{equation}
is the four vector of the matter polarization. In Eqs.~(\ref{eq:jms})
and~(\ref{eq:lambdams}), $n_{f}$ is the invariant number density
(the density in the rest frame of fermions), $\bm{\zeta}_{f}$ is
the invariant polarization (the polarization in the rest frame of
fermions), and $u_{f}^{\mu}=\left(u_{f}^{0},\mathbf{u}_{f}\right)$
is the four velocity. To derive Eqs.~(\ref{eq:flavnueffp})-(\ref{eq:lambdams})
it is crucial that background fermions have constant velocity. Only
in this situation one can make a boost to the rest frame of the fermions
where $n_{f}$ and $\bm{\zeta}_{f}$ are defined. The explicit form of the coefficients $q_{\alpha,f}^{(1,2)}$ in Eq.~(\ref{eq:flavnueffp})
can be found in Ref.~\citenum{DvoStu02}.

Nowadays it is experimentally confirmed
that the flavor neutrino eigenstates are the superposition of the
neutrino mass eigenstates, $\psi_{i}$, $i=1,2,\dotsc$,
\begin{equation}\label{eq:nupsi}
  \nu_{\alpha}=\sum_{i}U_{\alpha i}\psi_{i},
\end{equation}
where $\left(U_{\alpha i}\right)$ is the unitary mixing matrix. The
transformation in Eq.~(\ref{eq:nupsi}) diagonalizes the neutrino
mass matrix. Only using the neutrino mass eigenstates we can reveal
the nature of neutrinos, i.e. say whether they are Dirac or Majorana
particles. Despite the great experimental efforts to shed light upon the nature of neutrinos, this issue still remains open. Here we shall suppose that $\psi_{i}$ correspond to Dirac fields.

The effective Lagrangian for the interaction of $\psi_{i}$ with background
matter can be obtained using Eqs.~(\ref{eq:Lf}) and~(\ref{eq:nupsi}),
\begin{equation}\label{eq:Lg}
  \mathcal{L}_{\mathrm{eff}} =
  - \sum_{ij}
  \bar{\psi}_{i}\gamma_{\mu}^{\mathrm{L}}\psi_{j}\cdot g_{ij}^{\mu},
\end{equation}
where
\begin{equation}\label{eq:gab}
  g_{ij}^{\mu} = \sum_{\alpha}U_{\alpha i}^{*}U_{\alpha j}f_{\alpha}^{\mu},
\end{equation}
is the nondiagonal effective potential in the mass eigenstates basis.

Using Eq.~(\ref{eq:Lg}) one obtains that the corresponding Dirac
equations for the neutrino mass eigenstates are coupled,
\begin{equation}\label{eq:Depsi}
  \left[
    \mathrm{i}\gamma^{\mu}\partial_{\mu}-m_{i}-\gamma_{\mu}^{\mathrm{L}}g_{ii}^{\mu}
  \right]
  \psi_{i} =
  \sum_{j\neq i}\gamma_{\mu}^{\mathrm{L}}g_{ij}^{\mu}\psi_{j},
\end{equation}
where $m_{i}$ is the mass of $\psi_{i}$. One can proceed in the
analytical analysis of Eq.~(\ref{eq:Depsi}) if we exactly account
for only the diagonal effective potentials $g_{ii}^{\mu}$. To take
into account the r.h.s. of Eq.~(\ref{eq:Depsi}), depending on the
nondiagonal elements of the matrix $\left(g_{ij}^{\mu}\right)$, with
$i\neq j$, one should apply a perturbative method (see Sec.~\ref{sec:OSC}
below).

\section{Massive neutrinos in noninertial frames\label{sec:MASSNUNONIF}}

In this section we generalize the Dirac equation for a neutrino interacting
with a background matter to the situation when the velocity of the
matter motion is not constant. In particular, we study the case of
the matter rotation with a constant angular velocity. Then we obtain the
solution of the Dirac equation and find the energy spectrum.

If we discuss a neutrino mass eigenstate propagating in a nonuniformly
moving matter, the expressions for $f_{\alpha}^{\mu}$ in Eqs.~(\ref{eq:flavnueffp})-(\ref{eq:lambdams})
become invalid since they are derived under the assumption of the
unbroken Lorentz invariance. The most straightforward way to describe
the neutrino evolution in matter moving with an acceleration is to
rewrite the Dirac equation for a neutrino in the noninertial frame
where matter is at rest. In this case one can unambiguously define
the components of $f_{\alpha}^{\mu}$. Assuming that background fermions
are unpolarized, we find that in this reference frame
\begin{equation}\label{eq:f0nonin}
  f_{\alpha}^{0}=\sqrt{2}G_{\mathrm{F}}\sum_{f}q_{\alpha,f}^{(1)}n_{f}\neq0,
\end{equation}
with the rest of the effective potentials being equal to zero.

It is known that the motion of a test particle in a noninertial frame
is equivalent to the interaction of this particle with a gravitational
field. The Dirac equation for a massive neutrino moving in a curved space-time
and interacting with background matter can be obtained by the generalization
of Eq.~(\ref{eq:Depsi}) (see also Ref.~\citenum{GriMamMos80}),
\begin{equation}\label{eq:Depsicurv}
  \left[
    \mathrm{i}\gamma^{\mu}(x)\nabla_{\mu}-m
  \right]
  \psi =
  \frac{1}{2}\gamma_{\mu}(x)g^{\mu}
  \left[
    1-\gamma^{5}(x)
  \right]
  \psi,
\end{equation}
where $\gamma_{\mu}(x)$ are the coordinate dependent Dirac matrices,
$\nabla_{\mu}=\partial_{\mu}+\Gamma_{\mu}$ is the covariant derivative,
$\Gamma_{\mu}$ is the spin connection, $\gamma^{5}(x) = -\tfrac{\mathrm{i}}{4!} E^{\mu\nu\alpha\beta} \gamma_{\mu}(x) \gamma_{\nu}(x) \gamma_{\alpha}(x) \gamma_{\beta}(x)$,
$E^{\mu\nu\alpha\beta} = \tfrac{1}{\sqrt{-g}} \varepsilon^{\mu\nu\alpha\beta}$
is the covariant antisymmetric tensor in curved space-time, and $g=\det(g_{\mu\nu})$ is the determinant of the metric tensor $g_{\mu\nu}$. Note that in Eq.~(\ref{eq:Depsicurv})
we account for only the diagonal neutrino interaction with matter.
That is why we omit the index $i$ in order not to encumber the notation:
$m\equiv m_{i}$ etc. It should be noted that analogous Dirac equation
was discussed in Ref.~\citenum{PirRoyWud96}.

We shall be interested in the neutrino motion in matter rotating with
the constant angular velocity $\omega$. Choosing the corotating frame
we get that only $g^{0} \equiv g^{0}_{ii}$ is nonvanishing, cf. Eqs.~(\ref{eq:gab})
and~(\ref{eq:f0nonin}).

\subsection{Neutrino motion in a rotating frame\label{sub:ROTATION}}

The interval in the rotating frame is\cite{LanLif94}
\begin{equation}\label{eq:mertrot}
  \mathrm{d}s^{2} =
  g_{\mu\nu}\mathrm{d}x^{\mu}\mathrm{d}x^{\nu} =
  (1-\omega^{2}r^{2})\mathrm{d}t^{2} -
  \mathrm{d}r^{2}-2\omega r^{2}\mathrm{d}t\mathrm{d}\phi -
  r^{2}\mathrm{d}\phi^{2}-\mathrm{d}z^{2},
\end{equation}
where we use the cylindrical coordinates $x^{\mu}=(t,r,\phi,z)$. One can check that the metric tensor in Eq.~(\ref{eq:mertrot})
can be diagonalized, $\eta_{ab}=e_{a}^{\ \mu}e_{b}^{\ \nu}g_{\mu\nu}$,
if we use the following vierbein vectors:
\begin{align}\label{eq:vierb}
  e_{0}^{\ \mu}= &
  \left(
    \frac{1}{\sqrt{1-\omega^{2}r^{2}}},0,0,0
  \right),
  \nonumber
  \\
  e_{1}^{\ \mu}= & (0,1,0,0),
  \nonumber
  \\
  e_{2}^{\ \mu}= &
  \left(
    \frac{\omega r}{\sqrt{1-\omega^{2}r^{2}}},0,\frac{\sqrt{1-\omega^{2}r^{2}}}{r},0
  \right),
  \nonumber
  \\
  e_{3}^{\ \mu}= & (0,0,0,1).
\end{align}
Here $\eta_{ab}=\text{diag}(1,-1,-1,-1)$ is the metric in a locally
Minkowskian frame.

Let us introduce the Dirac matrices in a locally Minkowskian frame
by $\gamma^{\bar{a}}=e_{\ \mu}^{a}\gamma^{\mu}(x)$, where $e_{\ \mu}^{a}$
is the inverse vierbein: $e_{\ \mu}^{a}e_{b}^{\ \mu}=\delta_{b}^{a}$.
Starting from now, we shall mark an index with a bar to demonstrate that
a gamma matrix is defined in a locally Minkowskian frame.
As shown in Ref.~\citenum{Dvo14}, $\gamma^{5}(x) = \mathrm{i} \gamma^{\bar{0}} \gamma^{\bar{1}} \gamma^{\bar{2}} \gamma^{\bar{3}} = \gamma^{\bar{5}}$
does not depend on coordinates.

After the straightforward calculation of the spin connection on the basis of Eq.~(\ref{eq:vierb}), the Dirac Eq.~(\ref{eq:Depsicurv})
can be rewritten as
\begin{align}\label{eq:Derotf}
  [\mathcal{D} - & m]\psi =
  \frac{1}{2} \sqrt{1-\omega^{2}r^{2}}\gamma^{\bar{0}}g^{0}
  (1-\gamma^{\bar{5}})\psi,
  \nonumber
  \\
  \mathcal{D} = &
  \mathrm{i} \frac{\gamma^{\bar{0}}+\omega r\gamma^{\bar{2}}}{\sqrt{1-\omega^{2}r^{2}}}
  \partial_{0} +
  \mathrm{i} \gamma^{\bar{1}}
  \left(
    \partial_{r}+\frac{1}{2r}
  \right) +
  \mathrm{i} \gamma^{\bar{2}}
  \frac{\sqrt{1-\omega^{2}r^{2}}}{r}\partial_{\phi} +
  \mathrm{i} \gamma^{\bar{3}}\partial_{z}
  \notag
  \\
  & -
  \frac{\omega}{2(1-\omega^{2}r^{2})}
  \gamma^{\bar{3}}\gamma^{\bar{5}}.
\end{align}
The analogous Dirac equation was recently derived in Ref.~\citenum{Bak13}.
Since Eq.~(\ref{eq:Derotf}) does not explicitly contain $t$, $\phi$,
and $z$, its solution can be expressed as
\begin{equation}\label{eq:psitildepsi}
  \psi =
  \exp
  \left(
    -\mathrm{i}Et+\mathrm{i}J_{z}\phi+\mathrm{i}p_{z}z
  \right)
  \psi_{r},
\end{equation}
where $\psi_{r}=\psi_{r}(r)$ is the spinor depending on the radial
coordinate, $J_{z}=\tfrac{1}{2}-l$ (see, e.g., Ref.~\citenum{SchWieGre83}),
and $l=0,\pm1,\pm2,\dotsc$.

In Eq.~(\ref{eq:Derotf}) one can neglect terms $\sim(\omega r)^{2}$.
Indeed, if we study a neutrino in a rotating pulsar, then $r\lesssim10\thinspace\text{km}$
and $\omega\lesssim10^{3}\thinspace\text{s}^{-1}$. Thus $(\omega r)^{2}\lesssim1.1\times10^{-3}$
is a small parameter. Therefore Eq.~(\ref{eq:Derotf}) can be transformed
to
\begin{multline}\label{eq:Depsisimp}
  \bigg[
    \mathrm{i}\gamma^{\bar{1}}
    \left(
      \partial_{r}+\frac{1}{2r}
    \right) -
    \gamma^{\bar{2}}
    \left(
      \frac{J_{z}}{r}-\omega r E
    \right) +
    \gamma^{\bar{0}}
    \left(
      E-\frac{g^{0}}{2}
    \right) -
    \gamma^{\bar{3}}p_{z}
    \\
    + \frac{g^{0}}{2}\gamma^{\bar{0}}\gamma^{\bar{5}} -
    \frac{\omega}{2}\gamma^{\bar{3}}\gamma^{\bar{5}} - m
  \bigg]
  \psi_{r}=0,
\end{multline}
where we keep only the terms linear in $\omega$. It should be noted
that the term $\sim\omega\gamma^{\bar{3}}\gamma^{\bar{5}}$ in Eq.~(\ref{eq:Depsisimp})
is equivalent to the neutrino interaction with matter moving with
an effective velocity.

The solution of Eq.~\eqref{eq:Depsisimp} can be presented in the form\cite{Dvo14}, $\psi_{r}^{\mathrm{L}} = (0,\eta)^\mathrm{T}$ and $\psi_{r}^{\mathrm{R}} = (\xi,0)^\mathrm{T}$, where
\begin{equation}\label{eq:etaxi}
  \eta =
  \left(
    \begin{array}{c}
      -\mathrm{i}C_{1}I_{N,s} \\
      C_{2}I_{N-1,s}
    \end{array}
  \right),
  \quad
  \xi =
  \left(
    \begin{array}{c}
      C_{3}I_{N,s} \\
      -\mathrm{i}C_{4}I_{N-1,s}
    \end{array}
  \right).
\end{equation}
Here $N=0,1,2,\dotsc$, $s=N-l$,
$I_{N,s}=I_{N,s}(\rho)$ is the Laguerre function, and $\rho = E \omega r^2$. The explicit form of the Laguerre function can be found, e.g., in Ref.~\citenum{Dvo14}. To derive Eq.~\eqref{eq:etaxi} we use the Dirac matrices in the chiral representation\cite{ItzZub80}.

In the important case when $\omega\ll g^{0}$, the coefficients $C_i$, $i=1,\dots,4$, in Eq.~\eqref{eq:etaxi} are expressed in the following way\cite{Dvo14}:
\begin{align}\label{eq:C1-4slowrot}
  C_{1}^{2} \approx &
  \frac{E_{A}\omega}{2\pi}\frac{E_{A}-p_{z}-g^{0}}{E_{A}-g^{0}},
  \quad
  C_{3}^{2} \approx \frac{\omega}{2\pi}
  \left(
    E_{S}+p_{z}
  \right),
  \nonumber
  \\
  C_{2}^{2} \approx &
  \frac{E_{A}\omega}{2\pi}\frac{E_{A}+p_{z}-g^{0}}{E_{A}-g^{0}},
  \quad
  C_{4}^{2} \approx \frac{\omega}{2\pi}
  \left(
    E_{S}-p_{z}
  \right).
\end{align}
It should be noted that the solutions presented in Eqs.~(\ref{eq:etaxi}) and~(\ref{eq:C1-4slowrot})
satisfy the normalization condition,
\begin{equation}\label{eq:normgen}
  \int\psi_{N,s,p_{z}}^{\dagger}(x)
  \psi_{N',s',p'_{z}}(x)\sqrt{-g}\mathrm{d}^{3}x =
  \delta_{NN'}\delta_{ss'}
  \delta
  \left(
    p_{z}-p'_{z}
  \right).
\end{equation}
Here $\psi$ and $\psi_{r}$ are related by Eq.~(\ref{eq:psitildepsi}).

The energy levels in Eq.~\eqref{eq:C1-4slowrot} are
\begin{align}\label{eq:Energylev}
  \left[
    E_{A}-2N\omega-g^{0}
  \right]^{2} =
  & (2N\omega)^{2}+4N\omega g^{0} +
  \left(
    p_{z}-\frac{\omega}{2}
  \right)^{2},
  \nonumber
  \\
  \left[
    E_{S}-2N\omega
  \right]^{2} =
  & (2N\omega)^{2} +
  \left(
    p_{z}+\frac{\omega}{2}
  \right)^{2},
\end{align}
where $E_{A}$ and $E_{S}$ are the energies
of active and sterile neutrinos respectively. Comparing the expression
for $E_{S}\approx2N\omega+\sqrt{(2N\omega)^{2}+p_{z}^{2}}$ with the
energy of a neutrino in an inertial nonrotating frame $\sqrt{\mathbf{p}_{\perp}^{2}+p_{z}^{2}}$,
where $\mathbf{p}_{\perp}$ is the momentum in the equatorial plane,
we can identify $2N\omega$ inside the square root as $|\mathbf{p}_{\perp}|$.
It should be also noted that the term $2N\omega$, which additively
enters to both $E_{A}$ and $E_{S}$, is due to the noninertial effects
for a Dirac fermion in a rotating frame\cite{HehNi90}.

We can also get the corrections to
the energy levels due to the nonzero mass, $E_{A,S}\to E_{A,S}+E_{A,S}^{(1)}$.
On the basis of Eq.~(\ref{eq:etaxi}) and~(\ref{eq:C1-4slowrot}) one finds the
expression for $E_{A,S}^{(1)}$ in the limit $\omega\ll g^{0}$,
\begin{equation}\label{eq:EAS1}
  E_{A}^{(1)} =
  \frac{m^{2}}{2
  \left(
    E_{A}-2N\omega-g^{0}
  \right)
  },
  \quad
  E_{S}^{(1)} =
  \frac{m^{2}}{2
  \left(
    E_{S}-2N\omega
  \right)
  }.
\end{equation}
If we discuss neutrinos moving along the rotation axis, then $2N\omega\ll|p_{z}|$.
Using Eq.~(\ref{eq:Energylev}) we get the energy levels of active
neutrinos in this case
\begin{equation}\label{eq:Eadecom}
  E_{A} =
  |p_{z}|+g^{0}
  \left(
    1+\frac{2N\omega}{|p_{z}|}
  \right) +
  2N\omega +
  \frac{2(N\omega)^{2}}{|p_{z}|} +
  \frac{m^{2}}{2|p_{z}|},
\end{equation}
where we also keep the mass correction in Eq.~(\ref{eq:EAS1}). One
can see in Eq.~(\ref{eq:Eadecom}) that $|p_{z}|+g^{0}+\frac{m^{2}}{2|p_{z}|}$
corresponds to the energy of a left-handed neutrino interacting with background matter in a flat space-time. The rest
of the terms in Eq.~(\ref{eq:Eadecom}) are the corrections due to
the matter rotation.

\section{Flavor oscillations of Dirac neutrinos in rotating matter\label{sec:OSC}}

In this section we study the evolution of the system of massive mixed
neutrinos in rotating matter. We formulate the initial condition for
this system and derive the effective Schr\"{o}dinger equation which governs
neutrino flavor oscillations. Then we find the correction to the resonance
condition owing to the matter rotation and estimate its value for
a millisecond pulsar.

We can generalize the results of Sec.~\ref{sec:MASSNUNONIF} to include
different neutrino eigenstates. The interaction of neutrino mass eigenstates
with background matter is nondiagonal, cf. Eq.~(\ref{eq:Lg}). Therefore
the generalization of Eq.~(\ref{eq:Derotf}) for several mass eigenstates
$\psi_{i}$ reads
\begin{equation}\label{eq:Derotfpsia}
  \left[
    \mathcal{D}-m_{i}
  \right] \psi_{i} =
  \frac{1}{2}\gamma^{\bar{0}}
  g_{i}^{0} (1-\gamma^{\bar{5}}) \psi_{i} +
  \frac{1}{2}\gamma^{\bar{0}}
  \sum_{j\neq i}
  g_{ij}^{0}(1-\gamma^{\bar{5}})\psi_{j},
\end{equation}
where $g_{i}^{0}\equiv g_{ii}^{0}$ and $g_{ij}^{0}$ are the time
components of the matrix $\left(g_{ij}^{\mu}\right)$ given in Eq.~(\ref{eq:gab}),
$m_{i}$ is the mass of $\psi_{i}$, and $\mathcal{D}$ can be found
in Eq.~(\ref{eq:Derotf}). As in Sec.~\ref{sec:MASSNUNONIF}, we
omitted the term $(\omega r)^{2}\ll1$ in Eq.~(\ref{eq:Derotfpsia}).
Note that Eq.~(\ref{eq:Derotfpsia}) is a generalization of Eq.~(\ref{eq:Depsi})
for a system of the neutrino mass eigenstates moving in a rotating
frame.

We shall study the evolution of active ultrarelativistic neutrinos
and neglect neutrino-antineutrino transitions. In this case we can
restrict ourselves to the analysis of two component spinors. The general solution
of Eq.~(\ref{eq:Derotfpsia}) has the form,
\begin{equation}\label{eq:gensolosc}
  \eta_{i}(x) =
  \sum_{N,s}
  \int\frac{\mathrm{d}p_{z}}{\sqrt{2\pi}}
  a_{N,s,p_{z}}^{(i)}
  e^{\mathrm{i}p_{z}z+\mathrm{i}J_{z}\phi}
  u_{N,s,p_{z}}(r)
  e^{-\mathrm{i}E_{i}t},
\end{equation}
where $u_{N,s,p_{z}}$ are the basis spinors and $a_{N,s,p_{z}}^{(i)} = a_{N,s,p_{z}}^{(i)}(t)$ are the $c$-number functions. The energy levels $E_{i}$ are given in Eq.~(\ref{eq:Eadecom}) with
$m\to m_{i}$. Here we omit the subscript $A$ in order not to encumber
the notation. Our goal is to find the coefficient $a_{N,s,p_{z}}^{(i)}=a_{N,s,p_{z}}^{(i)}(t)$.
We neglect the small ratio $\omega/g_{i}^{0}$ in Eq.~(\ref{eq:gensolosc}).

Considering the system of two neutrino mass eigenstates, $i=1,2$, parameterized with one mixing angle $\theta$, and choosing the appropriate initial condition\cite{Dvo14}, on the basis
of Eq.~(\ref{eq:Derotfpsia}) we get the effective Schr\"{o}dinger equation
for $\tilde{\Psi}^{\mathrm{T}}=(a_{1},a_{2})$,%
\begin{equation}\label{eq:ScheqtildePsi}
  \mathrm{i}\frac{\mathrm{d}\tilde{\Psi}}{\mathrm{d}t} =
  \left(
    \begin{array}{cc}
      0 & g_{12}^{0}
      \exp
      \left[
        \mathrm{i}
          \left(E_{1}-E_{2}
        \right)
        t
      \right]
      \\
      g_{12}^{0}
      \exp
      \left[
        \mathrm{i}
        \left(
          E_{2}-E_{1}
        \right)
        t
      \right] & 0
    \end{array}
  \right)
  \tilde{\Psi}.
\end{equation}
Here we omitted all the indexes of $a_{i}$ besides $i=1,2$. It is
convenient to introduce the modified effective wave function $\Psi=\mathcal{U}_{3}\tilde{\Psi}$,
where $\mathcal{U}_{3}=\text{diag}\left(e^{\mathrm{i}\Omega t/2},e^{-\mathrm{i}\Omega t/2}\right)$,
$\Omega=E_{1}-E_{2}$. Using Eq.~(\ref{eq:ScheqtildePsi}), we get
for $\Psi$
\begin{equation}\label{eq:ScheqPsi}
  \mathrm{i}\frac{\mathrm{d}\Psi}{\mathrm{d}t} =
  \left(
    \begin{array}{cc}
      \Omega/2 & g_{12}^{0} \\
      g_{12}^{0} & -\Omega/2
    \end{array}
  \right)
  \Psi.
\end{equation}
Note that Eq.~(\ref{eq:ScheqPsi}) has the form of the effective
Schr\"{o}dinger equation one typically deals with in the study of neutrino
flavor oscillations in background matter.

If the transition probability for $\nu_{\alpha}\leftrightarrow\nu_{\beta}$
is close to one, i.e. $P_{\nu_{\beta}\to\nu_{\alpha}}=\left|\left\langle \nu_{\alpha}(t)|\nu_{\beta}(0)\right\rangle \right|^{2}\approx1$,
flavor oscillations of neutrinos are said to be at resonance. Using
Eqs.~(\ref{eq:nupsi}), (\ref{eq:f0nonin}), (\ref{eq:Eadecom}), and~\eqref{eq:ScheqPsi},
the resonance condition can be written as,
\begin{equation}\label{eq:rescondgen}
  \left(
    f_{\alpha}^{0}-f_{\beta}^{0}
  \right)
  \left(
    1+\frac{2N\omega}{|p_{z}|}
  \right) +
  \frac{\Delta m^{2}}{2|p_{z}|}\cos2\theta=0,
\end{equation}
where $\Delta m^{2}=m_{1}^{2}-m_{2}^{2}$ is the mass squared difference.

Let us consider electroneutral background matter composed of electrons,
protons, and neutrons. If we study the $\nu_{e}\to\nu_{\alpha}$ oscillation
channel, where $\alpha=\mu,\tau$,
we get that $f_{\nu_{\alpha}}^{0}=-\tfrac{1}{\sqrt{2}}G_{\mathrm{F}}n_{n}$
and $f_{\nu_{\beta}}^{0}\equiv f_{\nu_{e}}^{0}=\sqrt{2}G_{\mathrm{F}}\left(n_{e}-\tfrac{1}{2}n_{n}\right)$,
where $n_{e}$ and $n_{n}$ are the densities of electrons and neutrons.
Using Eq.~(\ref{eq:rescondgen}), we obtain that
\begin{equation}\label{eq:resnuenumu}
  \sqrt{2}G_{\mathrm{F}}n_{e}
  \left(
    1+\frac{2N\omega}{|p_{z}|}
  \right) =
  \frac{\Delta m^{2}}{2|p_{z}|}\cos2\theta.
\end{equation}
At the absence of rotation, $\omega=0$, Eq.~(\ref{eq:resnuenumu})
is equivalent to the Mikheyev-Smirnov-Wolfenstein resonance condition
in background matter\cite{BleSmi13}.

Let us evaluate the contribution of the matter rotation to the resonance
condition in Eq.~(\ref{eq:resnuenumu}) for a neutrino emitted inside
a rotating pulsar. We make a natural assumption that for a corotating
observer neutrinos are emitted in a spherically symmetric way
from a neutrinosphere. That is we should take that $l\approx0$ and
$N\approx s$. Then the trajectory of a neutrino is deflected because
of the noninertial effects and the interaction with background matter.
The radius $\mathcal{R}$ of the trajectory can be found from
\begin{equation}\label{eq:trajrad}
  \mathcal{R}^{2} =
  2|p_{z}|\omega
  \int_{0}^{\infty}
  r^{2}|u_{N,s,p_{z}}(r)|^{2}r\mathrm{d}r
  \approx
  \frac{2N}{|p_{z}|\omega},
\end{equation}
where we take into account that $N\gg1$.

We shall assume that $\mathcal{R}\sim R_{\mathrm{0}}$, where $R_{\mathrm{0}}=10\thinspace\text{km}$
is the pulsar radius. In this case neutrinos escape a pulsar. Taking
that $\omega=10^{3}\thinspace\text{s}^{-1}$ and using Eq.~(\ref{eq:trajrad}),
we get that the correction to the resonance condition in Eq.~(\ref{eq:resnuenumu})
is $\frac{2N\omega}{|p_{z}|}\approx\left(R_{\mathrm{0}}\omega\right)^{2}\approx10^{-3}$.
The obtained correction to the effective number density is small but
nonzero. This result corrects our previous statement\cite{DvoDib10}
that a matter rotation does not contribute neutrino flavor oscillations.

\section{Conclusion\label{sec:SUMMARY}}

In conclusion we notice that we have studied the evolution of massive
mixed neutrinos in nonuniformly moving background matter. The interaction
of neutrinos with background fermions is described in frames of the
Fermi theory (see Sec.~\ref{sec:NUMATFLS}). A particular case of the
matter rotating with a constant angular velocity has been studied
in Sec.~\ref{sub:ROTATION}. We have derived the Dirac equation for
a weakly interacting neutrino in a rotating frame and found its solution
in case of ultrarelativistic neutrinos, cf. Eqs.~(\ref{eq:etaxi}) and~(\ref{eq:C1-4slowrot}).
The energy spectrum obtained in Eqs.~(\ref{eq:Energylev}) and~(\ref{eq:EAS1})
includes the correction owing to the nonzero neutrino mass.

We have used the Dirac equation in a noninertial frame, cf. Eq.~(\ref{eq:Depsicurv}),
as a main tool for the study of the neutrino motion in matter moving
with an acceleration. To develop the quantum mechanical description
of such a neutrino we have chosen a noninertial frame where matter
is at rest. In this frame the effective potential of the neutrino-matter
interaction is well defined. However, the wave equation for a neutrino
turns out to be more complicated since one has to deal with noninertial
effects.

In Sec.~\ref{sec:OSC} we have generalized our results to include
various neutrino generations as well as mixing between them. We have
derived the effective Schr\"{o}dinger equation which governs neutrino
flavor oscillations. We have obtained the correction to the resonance
condition in background matter owing to the matter rotation. Studying
neutrino oscillations in a millisecond pulsar, we have obtained that
the effective number density changes by $0.1\thinspace\%$ owing to the matter rotation.

Despite the obtained correction is small, we may suggest that our
results can have some implication to the explanation of great linear
velocities of pulsars. It was suggested in Ref.~\citenum{KusSeg96}
that an asymmetry in neutrino oscillations in a magnetized pulsar
can explain a great linear velocity of the compact star. An evidence
for the alignment of the angular and the linear velocity vectors of pulsars
was reported in Ref.~\citenum{Joh05}. Therefore we may suggest that
neutrino flavor oscillations in a rapidly rotating pulsar can contribute
to its linear velocity. It should be noted that neutrino spin-flavor
oscillations, including noninertial effects, in a rapidly rotating
magnetized star were studied in Ref.~\citenum{Lam05} in the context
of the explanation of high linear velocities of pulsars.

Finally, we mention that the Dirac equation for a fermion, electroweakly interacting with the rotating background matter, was recently solved\cite{Dvo15}. The vierbein vectors, different from these in Eq.~\eqref{eq:vierb}, were used in Ref.~\citenum{Dvo15}. Comparing the energy levels obtained in Ref.~\citenum{Dvo15} with the results of the general analysis\cite{HehNi90}, one concludes that the vierbein used in Ref.~\citenum{Dvo15} is more appropriate for the description of ultrarelativistic particles like neutrinos.

\section*{Acknowledgments}

I am thankful to the organizers of the International Conference on Massive Neutrinos for the invitation and support, to S.~P.~Gavrilov for helpful comments, to FAPESP (Brazil) for the Grant No.~2011/50309-2,
to the Tomsk State University Competitiveness Improvement Program and to RFBR (research project No.~15-02-00293) for partial support.

\end{document}